\documentclass[reprint,
aps,superscriptaddress,twocolumn,showpacs,longbibliography]{revtex4-2}
\usepackage{graphicx,color}
\usepackage{amsmath,amsfonts,enumerate,amsthm,amssymb,bbm}
\usepackage[colorlinks=true,citecolor=blue,linkcolor=magenta]{hyperref}
\usepackage{multirow}
\usepackage{bbold}
\usepackage{braket}
\usepackage{soul}
\usepackage[caption = false]{subfig}
\usepackage{scrextend}
\usepackage{xcolor}
\usepackage{dsfont}
\usepackage{nicefrac}
\usepackage[linesnumbered,ruled,vlined]{algorithm2e}
\usepackage{tikz}
\usepackage{quantikz}
\usepackage{apptools}

\mathchardef\ordinarycolon\mathcode`\:
\mathcode`\:=\string"8000
\begingroup \catcode`\:=\active
  \gdef:{\mathrel{\mathop\ordinarycolon}}
\endgroup

\theoremstyle{plain}

\theoremstyle{definition}

\theoremstyle{remark}

\AtAppendix{\counterwithin{thm}{section}}
\AtAppendix{\counterwithin{corol}{section}}
\AtAppendix{\counterwithin{lem}{section}}
\AtAppendix{\counterwithin{propos}{section}}
\AtAppendix{\counterwithin{defn}{section}}
\AtAppendix{\counterwithin{rmk}{section}}
\AtAppendix{\counterwithin{ex}{section}}

\newcommand{\thv}{\vec{\theta}}

\renewcommand{\vec}[1]{\boldsymbol{#1}}

\renewcommand{\vec}[1]{\boldsymbol{#1}}

\begin{document}

\title{Learning Circuits with Infinite Tensor Networks}

\author{Joe Gibbs}
\affiliation{School of Mathematics and Physics, University of Surrey, Guildford, GU2 7XH, UK}
\affiliation{AWE, Aldermaston, Reading, RG7 4PR, UK}

\author{Lukasz Cincio}
\affiliation{Theoretical Division, Los Alamos National Laboratory, Los Alamos, NM 87545, USA}
\date{\today}

\begin{abstract}
Hamiltonian simulation on quantum computers is strongly constrained by gate counts, 
motivating techniques to reduce circuit depths. 
While tensor networks are natural competitors to quantum computers, we instead leverage them to support circuit design, with datasets of tensor networks enabling a unitary synthesis inspired by quantum machine learning.
For a target simulation in the thermodynamic limit, translation invariance is exploited to significantly reduce the optimization complexity, avoiding a scaling with system size. 
Our approach finds circuits to efficiently prepare ground states, and perform time evolution on both infinite and finite systems with substantially lower gate depths than conventional Trotterized methods.
In addition to reducing CNOT depths, we motivate similar utility for fault-tolerant quantum algorithms, with a demonstrated  $5.2\times$ reduction in $T$-count to realize $e^{-iHt}$. 
The key output of our approach is the optimized unit-cell of a translation invariant circuit. This provides an advantage for Hamiltonian simulation of finite, yet arbitrarily large, systems on real quantum computers.

\end{abstract}

\maketitle

\section{Introduction}

\begin{figure*}
\centering
\vspace{-5mm}
\includegraphics[width=1.0\textwidth]{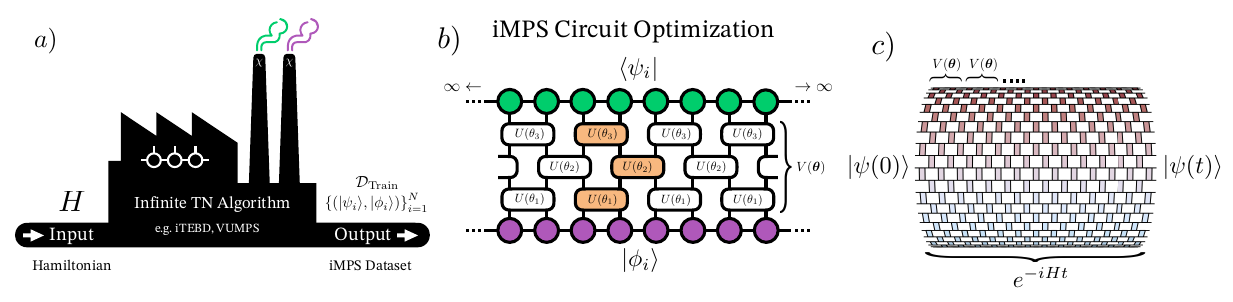}
\vspace{-7mm}
\caption{
\textbf{Overview. } 
a) Tensor network algorithms allow classical computation to support quantum simulation, creating datasets for a target Hamiltonian in the thermodynamic limit.  b) 
iMPS training states are used in variational unitary synthesis to learn efficient quantum circuits. Exploiting the translation invariance of the target system, a single 2-qubit unitary fully parameterizes an infinite layer, significantly reducing the optimization complexity. The output circuit is shorter in depth than equivalent circuits derived from Trotterization, and therefore more amenable to available quantum hardware. c) 
Optimized translation invariant layers are used for large-scale quantum simulation of finite systems, iterated to perform real-time evolution at circuit depths beyond classical simulability. 
}
\label{fig:Overview}
\end{figure*}

\definecolor{ketgreen}{HTML}{00CD6C}
\definecolor{brapurple}{HTML}{AF58BA}

\newcommand{\transfermatrix}{
\tikzset{every node/.style={minimum size=0.7cm, inner sep=2pt}}
\begin{tikzpicture}[baseline={(current bounding box.center)}]

  \coordinate (A1) at (0, 1);
  \coordinate (B1) at (1.1, 1);
  \coordinate (A2) at (0, 0);
  \coordinate (B2) at (1.1, 0);

  \node[draw, circle, fill=ketgreen, line width=1pt] (Atop) at (A1) {$A^\ast$};
  \node[draw, circle, fill=ketgreen, line width=1pt] (Btop) at (B1) {$B^\ast$};

  \node[draw, circle, fill=brapurple, line width=1pt] (Abot) at (A2) {$C$};
  \node[draw, circle, fill=brapurple, line width=1pt] (Bbot) at (B2) {$D$};

  \draw[line width=1pt] (Atop) -- (Abot);
  \draw[line width=1pt] (Btop) -- (Bbot);

  \draw[line width=1pt] (Atop) -- (Btop);
  \draw[line width=1pt] (Abot) -- (Bbot);

  \draw[line width=1pt] (Atop) -- ++(-0.75,0);
  \draw[line width=1pt] (Abot) -- ++(-0.75,0);
  \draw[line width=1pt] (Btop) -- ++(0.75,0);
  \draw[line width=1pt] (Bbot) -- ++(0.75,0);

\end{tikzpicture}
}

Quantum simulation remains the most promising application for quantum computers to demonstrate practical advantage over classical counterparts~\cite{feynman1982simulating}. Steady progress toward hardware with sufficient coherence times and qubit numbers has made such simulations increasingly feasible. 
However, high gate counts in target algorithms remains a primary obstacle. During the NISQ~\cite{preskill2018quantum} era, the accumulation of gate errors dominates and corrupts the signal. On future error-corrected quantum computers, this concern is replaced by unfavourable runtimes due to the overhead of magic-state distillation~\cite{bravyi2012magic}.
This creates a pressing need for algorithmic methods to reduce circuit depths, thereby improving the viability of quantum computations.

The ability to fully utilize classical computation will be crucial for enabling a near-term quantum advantage.
Factorization methods are increasingly applied to Hamiltonians to improve the efficiency of simulation on fault-tolerant hardware~\cite{lee2021even, caesura2025faster}.
Among the broad applications of machine learning, deep reinforcement learning has been used for gate count reduction~\cite{fosel2021quantum}, and $T$-count optimization~\cite{ruiz2025quantum}.
Representing quantum circuits as ZX-diagrams enabled the use of graph-based simplification rules to allow optimization of gate counts~\cite{duncan2020graph}.

Tensor networks~\cite{orus2019tensor} (TN) are a powerful framework for representing quantum many-body systems. The classical simulability of low-entanglement quantum states~\cite{Vidal2003Efficient} is exploited to find low-rank tensor descriptions of states and unitaries.
There has been growing interest in combining tensor network techniques with quantum circuit compilation~\cite{berezutskii2025tensor}, particularly for the construction of efficient quantum circuits to enhance quantum simulation. 
Approaches based on Matrix Product States (MPS) have been used to variationally compile circuits that prepare a target state~\cite{rudolph2022decomposition, rudolph2022synergy,dborin2022matrix,jamet2023anderson,rogerson2024quantum,lin2021real} or approximate $e^{-iHt}$~\cite{anselme2024combining,causer2023scalable,mc2023classically,gibbs2024deep,zhang2024scalable,le2025riemannian}, the unitary evolution under a Hamiltonian. 

Many physical systems of interest exhibit symmetries, such as translation invariance, that can be exploited to further reduce computational overhead. Mansuroglu \textit{et al} ~\cite{mansuroglu2023variational} used statevector simulation of small circuit patches of a translation invariant system and performed a variational optimization of a parameterized circuit. After upscaling to larger system sizes, they found lower errors for real-time dynamics compared to a Trotterization.
Gover \textit{et al}~\cite{gover2025fully} proposed a variational quantum algorithm with an ansatz inspired by infinite tensor network algorithms, to simulate a dynamical quantum phase transition in the infinite limit using a power-based approach. 
Our contribution is to similarly exploit translational invariance, while directly leveraging the power of infinite tensor networks using classical computation to support quantum computers.

Prior work in TN-enhanced quantum simulation has focused on finite systems, using MPS representations that scale in computational cost as $\mathcal{O}(L\chi^3)$, where $L$ is the system size and $\chi$ is the bond dimension. Exploiting translation invariance can remove this dependence on system size.
Therefore the goal of our work is the faithful optimization of the unit-cell of an infinite circuit to express and compress translation invariant unitaries.
This unit-cell can then be scaled up to arbitrary (but finite) size quantum simulations.
This computation is made possible through an optimization based on infinite Matrix Product States (iMPS)~\cite{vidal2007classical}, a framework originally introduced by Vidal, and later extended with improved convergence and efficiency
~\cite{zauner2018variational}.

We propose to use local fidelity as the basis for the cost function, which provides a unified approach to learn circuits for ground state preparation, time-evolved states, and unitary synthesis.
When learning a circuit to prepare a particular state, the error can be measured directly. This is in contrast with unitary compilation. Ideally, one would like to avoid complexities related to working with the full unitary using infinite Matrix Product Operators. To address that problem and associated increased computational costs, we use an approach inspired by quantum machine learning (QML) to perform the circuit compilation~\cite{caro2022outofdistribution}.

Our work opens a new direction in circuit depth optimization: compiling directly in the infinite limit. The approach offers scalable framework that avoid finite-size effects, enabling more accurate implementations of algorithmic primitives relevant to condensed matter and materials science.

\medskip

The rest of the paper is summarized as follows. Section~\ref{sec:methods} describes our representation of translation invariant circuits using iMPS, and motivates the methods used to enable a variational optimization.
In Section~\ref{sec:ground_state}, we first compile a circuit to approximately prepare a target ground state represented by an iMPS. We study the scaling of achievable error when preparing ground states of the Transverse-Field Ising Model (TFIM) Hamiltonian as the critical point is approached.
In Section~\ref{sec:state_evolution} we compile a short-depth circuit to prepare an initial state evolved under two target Hamiltonians (Thirring Model and 1D XXZ model). Compared to Trotterizations of the time-evolution operator, the optimized circuits prepare the target state to a given approximation error with shorter depths.

In Section~\ref{sec:unitary_compression} we generalize beyond state compilation, and demonstrate the QML-based compilation of the full unitary $e^{-iHt}$ for the TFIM and Thirring Hamiltonians. Our circuits find a lower approximation error to all equivalent depth Trotterizations.
To motivate a utility in the fault-tolerant (FT) regime of quantum simulations, we demonstrate in Section~\ref{sec:T-Count} the decomposition of our iMPS-compressed circuits to the Clifford+$T$ gate set can significantly reduce the $T$-count. Compared to circuits derived from Trotterization, our optimized circuits realize the unitary $e^{-iHt}$ with up to a $5.2\times$ reduction in $T$ gates to reach a target error.
Finally, in Section~\ref{sec:InfiniteToFinit}, we validate that our circuits compiled in the infinite limit successfully gain an advantage on finite qubit numbers, with state vector simulations on up to 30 qubits.

\medskip

Together, our results provide a scalable, symmetry-aware framework for quantum circuit compilation that leverages the structure of physical systems and the strengths of infinite tensor network methods. Our approach, summarized in Fig.~\ref{fig:Overview}, could play an important role in enabling practical quantum simulations through reduced algorithmic resources in both the NISQ and fault-tolerant regimes.

Note: While preparing our manuscript, we became aware of the recent work by Sokolov and Dziarmaga~\cite{sokolov2025bang}, who use infinite tensor networks through a bang-bang algorithm to variationally prepare ground states in one and two dimensions. 

\begin{figure*}[t!]
\centering
\vspace{-5mm}
\includegraphics[width=0.8\textwidth]{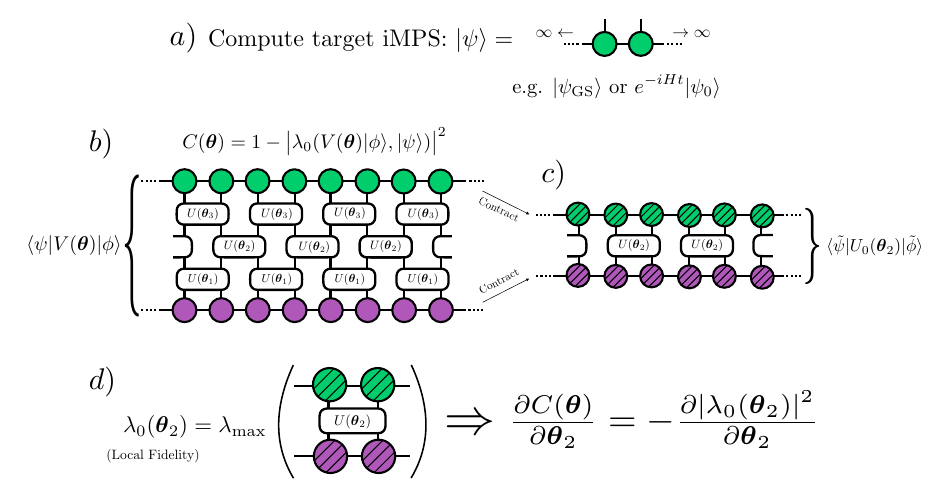}
\vspace{-3mm}
\caption{
\textbf{Contraction Illustration.}  Description of gradient calculation for a layer of the infinite parameterized circuit.
a) The target state, $|\psi\rangle$, is computed as an iMPS using an infinite tensor network algorithm. b) The cost function here is the local infidelity between $|\psi\rangle$, and the variational state $V(\boldsymbol{\theta})|\phi\rangle$,  involving operations on the mixed transfer matrix. c) We can isolate the contribution of the cost function with respect to the $2^\text{nd}$ layer, parameterized by $\thv_2$, by contracting the network above and below this layer. The stripes highlight the updated tensors due to layer contraction, representing the ``environment iMPS'' $|\tilde{\psi}\rangle$ and $|\tilde{\phi}\rangle$ as defined by Eq.~\eqref{eq:psi_tilde}. d) Then the local infidelity (written as a function of $\thv_2$) is given by the largest magnitude eigenvalue of the shown tensor network. This allows a partial derivative with respect to the cost function to be computed using a numerical finite-difference gradient.
}
\label{fig:Contract}
\end{figure*}

\section{Methods}\label{sec:methods}

\subsection{Infinite Matrix Product States}\label{sec:iMPS}

Infinite MPS~\cite{vidal2007classical} were introduced as a tensor network ansatz for describing translation invariant, infinite one-dimensional (1D) quantum states.
Throughout this work, we focus on compiling translation invariant states and unitaries with a 2-site unit cell, however our method can be easily extended to larger unit cells. 

\subsection{Calculating Fidelity}\label{sec:LocalFidelity}

Within our variational compilation, the efficient calculation of fidelity between states is an essential ingredient. Due to the orthogonality catastrophe in exponentially large Hilbert spaces, the global fidelity does not provide useful information on infinite systems, and in fact takes a binary value of 0 or 1 for normalized states. Therefore, we instead compute a local measure of fidelity per qubit. 

In iMPS calculations, the transfer matrix plays a central role. Quantum states represented by iMPS are normalized by scaling the leading eigenvalue of the transfer matrix to have magnitude 1.
Similarly, the local fidelity between two normalized iMPS $|\psi_1\rangle$ and $|\psi_2\rangle$ is given by the leading eigenvalue of the corresponding mixed transfer matrix:

If $|\psi_1\rangle$ and $|\psi_2\rangle$ are described by the iMPS site tensors $\{A,B\}$ and $\{C,D\}$ respectively, the following tensor network gives their mixed transfer matrix.
\begin{center}
\transfermatrix    
\end{center}
Consequently, the largest magnitude eigenvalue of this matrix gives the local fidelity between the two states, which we denote by
\begin{equation}
    f_\text{local} = \lambda_0(|\psi_1\rangle, |\psi_2 \rangle).
\end{equation}
Throughout this work we ensure all iMPS remain normalized, by imposing that $\big|\lambda_0(|\psi\rangle, |\psi \rangle) \big| = 1$.

To enable an efficient calculation of this leading eigenvalue, we use the implicitly restarted Arnoldi method, accessed via the `Arpack.jl' package~\cite{lehoucq1998arpack}. The iterative method targets the dominant eigenvalue and does not require the matrix is explicitly constructed, only its action when multiplied onto a vector. Therefore, the memory is dominated by the outputted vector after the Arnoldi iteration, which is $\mathcal{O}(\chi^2)$ as desired. 

\subsection{Gates}\label{sec:gates}
Our parameterized circuits are described by translation invariant layers of 2-qubit gates. The most expressive layer design is composed of $SU(4)$ gates.  Ignoring the global phase, these can be fully described by 15 parameters.
\begin{equation} \label{eq:SU(4)gate}
    U(\thv) = \text{exp}\left(-i \sum_{k=1}^{15} \thv_k G_k\right) \ ,   
\end{equation}
where the generator basis is constructed with 2-qubit Pauli matrices, minus the identity element, with $G = \{I,X,Y,Z\}^{\otimes 2} \setminus \{I \otimes I\}$. 
Alternate gate constructions could be chosen depending on the target problem, for example to exploit conserved quantities such as parity or particle number. One can also consider a gate choice with explicit decomposition designed to minimize the number of CNOT gates. In Sec.~\ref{sec:T-Count}, we use a gate construction designed to minimize the number of $R_z$ gates to aid the decomposition to a fault-tolerant gate set. Throughout the rest of our work we use $SU(4)$ layers defined in Eq~\eqref{eq:SU(4)gate}.

As we restrict ourselves to iMPS with 2-site unit cells, we introduce notation to distinguish the two configurations for infinite translation invariant layers of gates, acting on ``odd'' or ``even'' pairs of qubits.
A unitary with a subscript denotes an infinite layer of commuting gates defined by
\begin{equation*}
\begin{split}
    U_0 &= \bigotimes_{r \in \mathbb{Z}} U^{[2r, 2r+1]} \\
    U_1 &= \bigotimes_{r \in \mathbb{Z}} U^{[2r-1, 2r]}.    
\end{split}
\end{equation*}

All circuits considered in this work are then described by $V(\thv) = \prod_{l} U_{l\%2}(\thv_l)$, where ``$l\%2$'' is a shorthand for ``$l$ modulo 2''. 

The infinite time-evolving block decimation (iTEBD) algorithm~\cite{vidal2007classical} gives a fast and efficient method to apply an infinite and translation invariant layer of 2-qubit gates to an iMPS. 
This contraction is the only time we require explicit access to the bond matrices used in the iTEBD algorithm, and otherwise they are absorbed into the site tensors. 

\subsection{Optimization}\label{sec:Optimization}

The goal of our compilation is to optimize the parameterized infinite quantum circuit $V(\thv)$, to maximize the overlap between the set of states $\{V(\thv)|\phi_i \rangle \}_{i=1}^{N}$ and $\{|\psi_i \rangle \}_{i=1}^{N}$. 
$\{|\phi_i \rangle \}_{i=1}^{N}$ and $\{|\psi_i \rangle \}_{i=1}^{N}$ are both datasets of iMPS computed with an infinite tensor network algorithm, and depend on the target application.

Then, the most general cost function we consider has the form
\begin{equation}\label{eq:Cost_General}
    C(\thv) = 1 - \frac{1}{N}
    \sum_{i=1}^{N} \big|\lambda_0(V(\thv)  |\phi_i \rangle, | \psi_i \rangle)\big|^2.
\end{equation}

In the following we describe the computation of gradients of the cost function. For simplicity, in this section we set $N = 1$, however by the linearity of the gradient calculation it is trivially extended to the general form of Eq.~\eqref{eq:Cost_General}.

The key component in our cost function is the local fidelity between two states, and we use gradient-based methods to optimize this quantity. We target a global optimization, where the gradients for each $\thv_l$ defining the $l^\text{th}$ layer of gates are computed, and then updated simultaneously. To calculate the gradient vector $\frac{\partial C(\thv)}{\partial \thv_l}$, we first isolate the contribution to the cost function arising from the $l^{\mathrm{th}}$ unitary layer $U_{l\%2}(\thv_l)$,
by contracting the network above ($k > l$) and below ($k < l$) this layer,
given by

\begin{align*}
\langle\psi|V(\theta)|\phi\rangle &= \langle\psi|(\prod_l U_{l \% 2}(\theta_l))|\phi\rangle \\ &= \langle\psi|(\prod_{k>l} U_{k \% 2}(\theta_k)) U_{l\%2}(\theta_l) (\prod_{k<l} U_{k \% 2}(\theta_k))|\phi\rangle\\ & = \langle \tilde{\psi}| U_{l\%2}(\theta_l) |\tilde{\phi}\rangle \ ,
\end{align*}
where 
\begin{equation}\label{eq:psi_tilde}
\begin{split}
    |\tilde{\phi}\rangle &= \prod_{k < l} U_{k\%2}(\thv_k) |\phi\rangle \newline \\
    |\tilde{\psi}\rangle &= (\prod_{k > l} U_{k\%2}(\thv_k))^{\dagger} |\psi\rangle \ .
\end{split}
\end{equation}

This contraction is also shown graphically in Fig.~\ref{fig:Contract}~b) and~c). The cost function with respect to $\thv_l$ is then given by 

\begin{equation}
    C(\thv_l) = 1 - \big|\lambda_0 (U_{l\%2}(\thv_l) |\tilde{\phi} \rangle, | \tilde{\psi} \rangle) \big|^2 \ .
\end{equation}

The above cost function is computed from the largest magnitude eigenvalue of the tensor network shown in Fig.~\ref{fig:Contract}~d), and simplifies the gradient calculation as 
$$\frac{\partial C(\thv)}{\partial \thv_l} = \frac{\partial C(\thv_l)}{\partial \thv_l}.$$

Finally, the vector $\frac{\partial C(\thv_l)}{\partial \thv_l}$ is computed using a finite difference gradient calculation from the quantity shown in Fig.~\ref{fig:Contract} d). 

\begin{figure*}[t!]
\centering
\vspace{-5mm}
\includegraphics[width=0.8\textwidth]{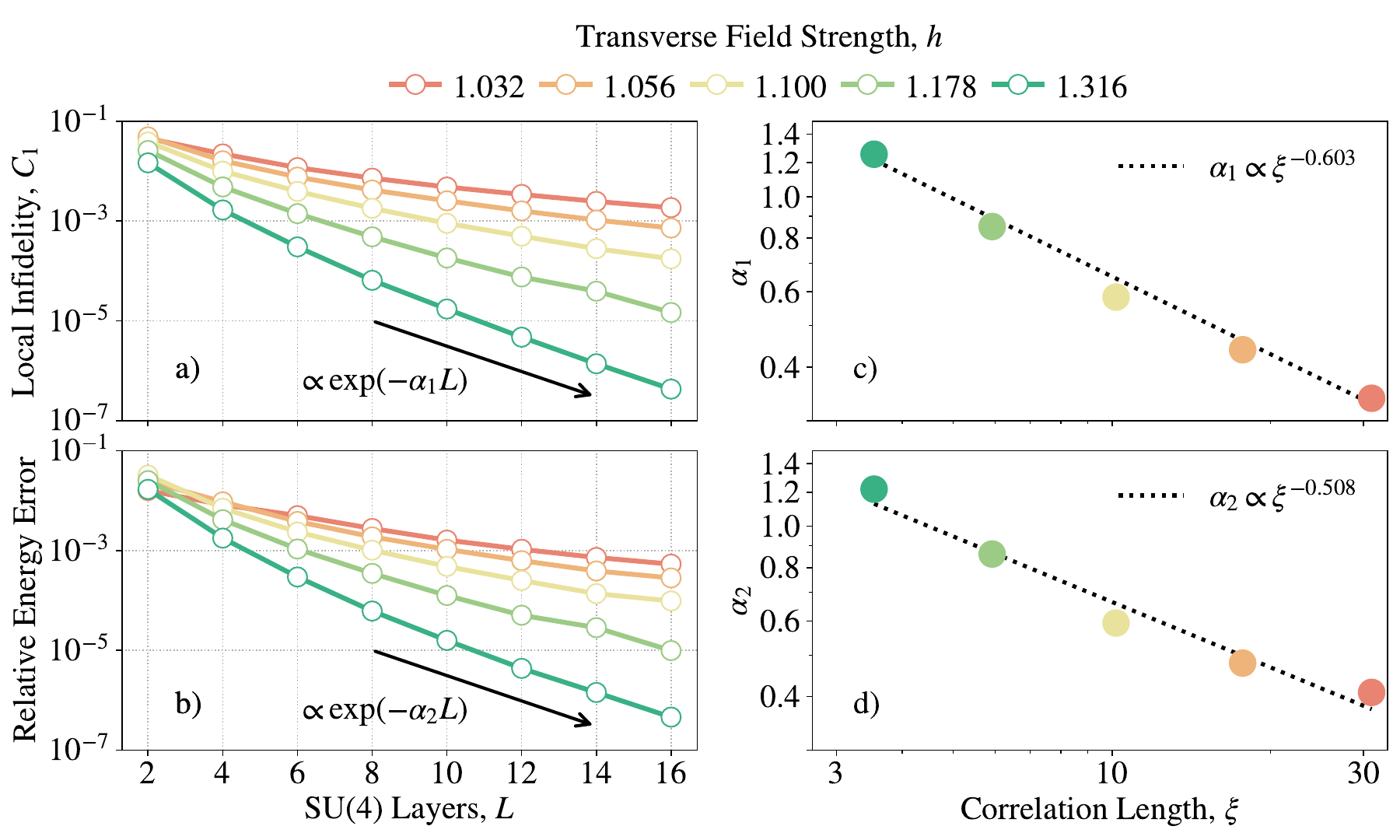}
\vspace{-3mm}
\caption{
\textbf{Circuit Compilation: Ground States. }
a) We apply our iMPS-based compilation to ground state preparation of the infinite 1D TFIM Hamiltonian (Eq.~\eqref{eq:Ham_TFIM}) for varying transverse field strengths $h$. The target ground states were computed as iMPS using iTEBD, with a maximum bond dimension of 128. The depths of the variational circuits are measured in number of layers of infinite translation invariant layers of $SU(4)$ gates, shown on the x-axis. The optimization uses the local infidelity cost function (Eq.~\eqref{eq:Cost_GS}) versus the target iMPS, and the final value upon convergence is displayed. b) To verify the cost function is faithful, we compute the corresponding relative error $|E - E_{\text{GS}}|/|E_{\text{GS}}|$ in the variational energy density, $E$, compared to the ground state energy density, $E_{\text{GS}}$. c-d) An exponential decay in both the local infidelity, and relative energy density error, as a function of circuit depth, $L$, is observed. Here, $\alpha_1$ and $\alpha_2$ denote the decay coefficients. Those coefficients vary as a function of the correlation length, $\xi$ of the target ground state, which a polynomial fit finds to be $\alpha_1 \propto \xi^{-0.603}$ and $\alpha_2 \propto \xi^{-0.508}$.
}
\label{fig:GroundStatePrep}

\end{figure*}

Given these gradient calculations are based on an iterative eigensolver, we can further optimize our computations. Using finite-difference gradients requires cost function values on small perturbations in the current parameter vector. Each of these involves an Arnoldi Iteration computation to determine the leading magnitude eigenvalue of a mixed Transfer Matrix. Iterative eigensolvers typically start with a random input state, which is refined towards the target eigenstate. However, this calculation can be significantly accelerated if an input vector close to the target eigenvector is inputted. 
    
In the finite-difference calculation we compute a reference cost function value $C(\thv_l)$ and 15 (for $SU(4)$ gates) perturbed cost function values $\{C(\thv_l + h \delta_{l,k})\}_{k=1}^{15}$. As we use a small step-size of $h=10^{-7}$, the corresponding mixed transfer matrix for each instance is very similar. Therefore, we save the corresponding eigenvector from the transfer matrix of the reference cost function and give this as input to the Arnoldi Iteration for the perturbed cost function calculations. This avoids redundant computation on similar matrices to accelerate the gradient calculation.

During the gradient calculation, the states $|\tilde{\psi}\rangle$ and $|\tilde{\phi}\rangle$ act as ``environment iMPS'', as they contain all information about the infinite tensor network above and below the $l^\text{th}$ infinite gate layer. We note that these environment iMPS can also be reused to avoid redundant computation. When moving to the gradient calculation on the $(l+1)^\text{th}$ layer, only a single infinite gate layer is applied with simple update to each environment iMPS
\begin{equation}\label{eq:psi_tilde}
\begin{split}
    |\tilde{\psi}\rangle &\rightarrow U_{(l+1)\%2}(\thv_{l+1})^{\dagger}|\tilde{\psi}\rangle \newline \\
    |\tilde{\phi}\rangle &\rightarrow U_{l\%2}(\thv_l)|\tilde{\phi}\rangle \ .
\end{split}
\end{equation}

The numerical optimizer used throughout our work is the gradient-based L-BFGS algorithm, provided by the ``Optim.jl'' package~\cite{mogensen2018optim}. To accelerate the calculation of gradients, we give as input a custom gradient function using the optimizations described above. 

\section{Results}\label{sec:results}

We perform a set of numerical experiments validating the utility of the iMPS-based circuit compilation for a~range of quantum simulation applications. We work with three Hamiltonians, which we detail here first.

\bigskip
1) The 1D Transverse-Field Ising model (TFIM), defined by
\begin{equation}\label{eq:Ham_TFIM}
    H = -J\sum_{i \in \mathbb{Z}} \sigma^z_i \sigma^z_{i+1} - h\sum_{i\in \mathbb{Z}} \sigma^x_i .
\end{equation}
Throughout our analysis we fix $J=1.0$.

\bigskip

2) The massive Thirring model~\cite{thirring1958soluble} is a theoretical model of a (1+1)D quantum field theory. After a lattice discretization in the Kogut-Susskind staggered formulation~\cite{kogut1975hamiltonian}, it can be represented by the following spin-$\frac12$ Hamiltonian
\begin{equation}\label{eq:Ham_Thirring}
\begin{split}
    H = \frac{i}{2a} &\sum_{i \in \mathbb{Z}} (\sigma^{-}_{i+1} \sigma^{+}_{i} - \sigma^{-}_{i} \sigma^{+}_{i+1}) \\ 
    + \frac{m}{2} &\sum_{i \in \mathbb{Z}} (-1)^i (\mathbb{1}-\sigma^z_i) \\ 
    + \frac{g}{4a} &\sum_{i \in \mathbb{Z}} (\mathbb{1}-\sigma^z_i) (\mathbb{1}-\sigma^z_{i+1}) \ . \\
\end{split}
\end{equation}
Throughout our analysis we fix $m=0.8$ and $g=0.4$.

\bigskip

3) The 1D XXZ model, defined by
\begin{equation}\label{eq:Ham_XXZ}
    H = \sum_{i \in \mathbb{Z}} J_x(\sigma^{x}_{i} \sigma^{x}_{i+1} + \sigma^{y}_{i} \sigma^{y}_{i+1}) + J_z \sigma^{z}_{i} \sigma^{z}_{i+1} \ .
\end{equation}

Throughout our analysis we fix $J_x=1.0$ and $J_z=0.5$.

\subsection{Ground-State Preparation}\label{sec:ground_state}

Our first demonstration is a circuit compilation to prepare the ground state of the Transverse-Field Ising Hamiltonian, defined in Eq.~\eqref{eq:Ham_TFIM}.

The system undergoes a phase transition from ferromagnetic to paramagnetic at $h=1$. At that value of $h$ the correlation length diverges. We explore the achievable fidelity with the exact ground state for a range of variational circuit depths as the field strength approaches this critical value. 

We work with the following cost function that minimizes the local infidelity between the ground state $|\psi_{\mathrm{GS}}\rangle$ and the variational state $V(\thv)  |\bf0 \rangle$:
\begin{equation}\label{eq:Cost_GS}
    C_{1}(\thv) = 1 - \big|\lambda_0(V(\thv)  |\boldsymbol{0} \rangle, | \psi_{\mathrm{GS}} \rangle)\big|^2 \ ,
\end{equation}
where $|\boldsymbol{0}\rangle = |0\rangle^{\otimes \infty}$.

The ``exact'' ground state is computed as an iMPS, $|\psi_{\mathrm{GS}}\rangle$, using iTEBD~\cite{vidal2007classical}. We perform imaginary time evolution using a high-accuracy Trotterization of the operator $e^{-\tau H}$, decreasing the imaginary time step-size $\Delta \tau$ until the energy density converges. We use a maximum bond dimension of 128, which for all field strengths allows the energy expectation to have a relative error less than $10^{-8}$ from the analytic ground-state energy value~\cite{pfeuty1970one}. 
For our ground state compilations, over a range of transverse-field strengths $h$, and variational circuit depths, we optimize the cost function defined in Eq.~\eqref{eq:Cost_GS} using the gradient-based method outlined in Sec.~\ref{sec:Optimization}. We use a simple identity initialization for our variational circuits. 

\begin{figure*}[t!]
\centering
\vspace{-5mm}
\includegraphics[width=0.9\textwidth]{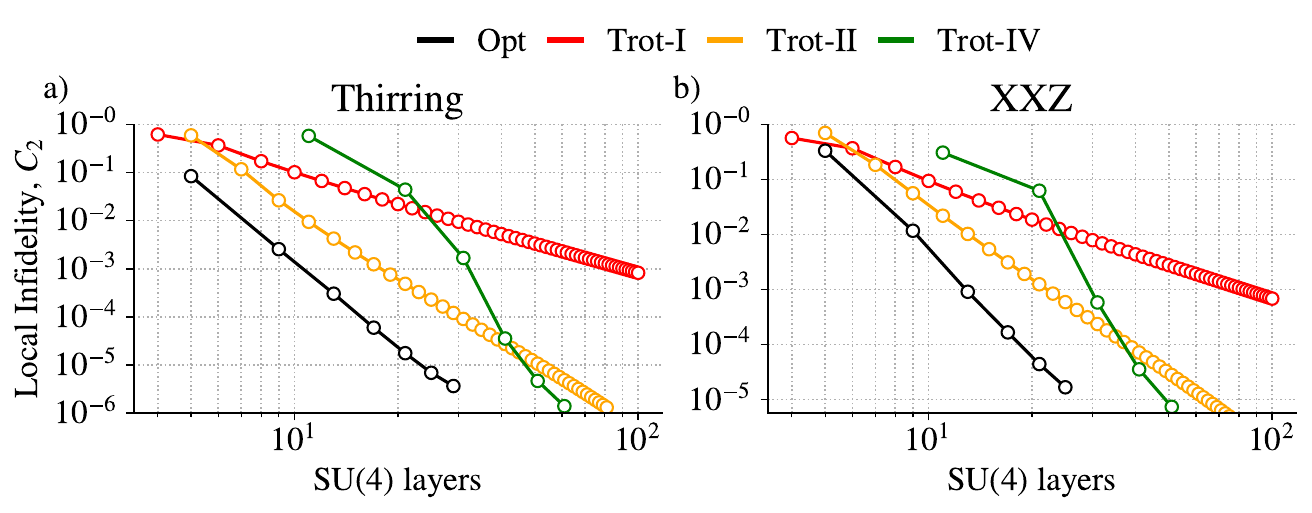}
\vspace{-3mm}
\caption{
\textbf{Circuit Compilation: Real-Time State Evolution.}
The target state is an initial state evolved under a Hamiltonian $H$, $|\psi_0\rangle =  e^{-iHt}|\phi_0\rangle$. 
For the demonstration with the Thirring (XXZ) model as the Hamiltonian, the initial state is the Néel state $|\phi_0\rangle = |10\rangle^{\otimes \infty}$ and is evolved for total time $t=8.0$ ($t=1.5$). The red/orange/green lines respectively correspond to 1st/2nd/4th-order Trotterizations. The local infidelity of the initial state evolved by increasing depth Trotterizations is computed with respect to the target state $|\psi_0\rangle$. 
The black curve shows the data from the iMPS-optimized circuits, produced by optimizing Eq.~\eqref{eq:Cost_State} to minimize the local infidelity of the variational state with respect to the target state. 
The Trotterized and iMPS-optimized circuits are compared in the local infidelity achieved to the target state, as a function of infinite layers of $SU(4)$ gates in the circuit.
For each Hamiltonian, we achieve over an order of magnitude reduction in error compared to equivalent depth Trotterization.}
\label{fig:StateEvol}
\end{figure*}

The phase diagram of the 1D TFIM has a critical point at $h=1$, with a corresponding infinite correlation length, where a finite depth circuit of nearest-neighbor gates will be unable to exactly prepare this state. Compared to the very accurate ground state expressed as an iMPS, we explore the rate of convergence as a function of circuit depth in both the local infidelity and the relative energy density error, and analyze how this changes as the critical point is approached.

Our numerical results are shown in Fig.~\ref{fig:GroundStatePrep}. We study the performance of the compilation for ground states of the TFIM with transverse-field strengths 
$h \in \{1.032,1.056,1.100,1,178,1.316\}$, with variational circuit depths up to 16 infinite layers of $SU(4)$ gates. 
We evaluate the performance of the optimization using two metrics: local infidelity and relative error in the energy per site. Our optimization directly minimizes the local infidelity via Eq.~\eqref{eq:Cost_GS}, and upon convergence we plot the final local infidelity value in Fig.~\ref{fig:GroundStatePrep} a). To verify this cost function is faithful, we also compute the corresponding variational energy density, $E$, and plot the relative energy density error, $|E -E_{\text{GS}}| / |E_{\text{GS}}|$ in Fig.~\ref{fig:GroundStatePrep} b).

For both metrics, we observe a smooth decrease in error as the variational circuit depth is increased, and the decrease is faster for target ground states with shorter correlation lengths. For the largest field strength studied ($h=1.316$) and deepest circuit (16 $SU(4)$ layers) we find both local infidelities and relative energy errors below $10^{-6}$.

As a function of circuit depth $L$, there is an observed exponential decay in both the local infidelity and relative energy error of the variational circuit compared to the target ground state. The rate of decay is dependent on the transverse field strength. The further from the critical point the ground state is, the faster the two error metrics decrease with circuit depth. We numerically quantify this by extracting the decay with an exponential fit for each data curve in Fig.~\ref{fig:GroundStatePrep} a) and b). The fit is performed for data points with $L\geq8$ as the exponential decay behavior is more clear for the deeper circuits. We can numerically quantify this by computing the average mean squared error (MSE) between the logarithm of the data points versus a linear fit with and without the $L<8$ data points. We find that for the local infidelity (energy error) the average MSE is reduced from 0.077 (0.10) to 0.0026 (0.0074), resulting in a reduction in error by a factor of 29.1 (14.0). Denoting the exponential decay constant for the local infidelity (energy error) by $\alpha_1$ ($\alpha_2$), we quantify how this quantity changes as the critical point is approached. In Fig.~\ref{fig:GroundStatePrep} c) and d), we plot the $\alpha$ values as a function of the correlation length $\xi$, of the target ground state. 
The correlation length is computed numerically using the relation $ \xi = -\frac{1}{\text{log}(\lambda_1)}$~\cite{schollwock2011density}, where $\lambda_1$ is the second largest magnitude eigenvalue of the ground state transfer matrix.
We find a polynomial relation between $\alpha$ and $\xi$, and a fit determines this to be approximately $\alpha_1 \propto \xi^{-0.603}$ and $\alpha_2 \propto \xi^{-0.508}$.

\subsection{Real-Time State Evolution}\label{sec:state_evolution}

Our second demonstration is the compilation of a circuit to prepare a state evolved under a given, local Hamiltonian. We use the following cost function:

\begin{equation}\label{eq:Cost_State}
    C_{2}(\thv) = 1 - \big|\lambda_0(V(\thv)  |\phi_0 \rangle, e^{-iHt}|\phi_{0} \rangle)\big|^2.
\end{equation}

In this demonstration, we choose the Thirring model defined in Eq.~$\eqref{eq:Ham_Thirring}$ and XXZ model in Eq.~$\eqref{eq:Ham_XXZ}$ as the Hamiltonians. The initial state used in  both calculations is the Néel state $|\phi_0\rangle = |10\rangle^{\otimes \infty}$, which we evolve up to $t=8.0$ and $t=1.5$ respectively. iTEBD is used to compute the target iMPS $e^{-iHt}|\phi_{0} \rangle$ with a maximum bond dimension of 128,
using a high-accuracy Trotterization converged in Trotter error well below the approximation errors introduced from compilation.

The goal here is to compile a circuit that prepares a target state with a specified fidelity, using shorter depth (i.e., fewer two-qubit gates) than a Trotterized evolution of the initial state. Variational circuits with infinite $SU(4)$ layers use an equivalent depth 2nd-order Trotterization as an initialization, followed by minimization of the cost function Eq.~\eqref{eq:Cost_State}. For a Trotterization $U(t)$ of the time evolution unitary, $e^{-iHt}$, accuracy is increased with repetitions over smaller timesteps, resulting in circuits of the form $U(t/k)^k$, where $k$ is the `Trotter Number'. We test our optimized circuits against 1st/2nd/4th-order Trotterizations, each over a wide range of Trotter numbers. The depths reported take account of simplifications of neighboring gate layers acting on the same pairs of qubits, resulting in all circuits comprising of alternating ``odd'' and ``even'' gate layers.

For all circuit depths, the variational circuits produced by our iMPS-based optimization find lower errors compared to an equivalent depth Trotterization of all orders, with the results shown in Fig.~\ref{fig:StateEvol}. For the Thirring (XXZ) Hamiltonian, at the deepest tested circuit depth of 29 (25) $SU(4)$ layers, we find error reductions by a factor of 32.3 (35.7).

Any $SU(4)$ unitary can be decomposed into at most 3 CNOT gates by the KAK decomposition~\cite{tucci2005introduction}. As a consequence, these results directly show the iMPS-optimized circuits can prepare the time-evolved state on a quantum computer to a target error rate with fewer CNOT gates per qubit compared to Trotterized circuits. As the error rate due to native 2-qubit gates is the dominant source of error on NISQ hardware, our results show positive evidence for an ability to support near-term quantum advantage. 

\begin{figure*}[t!]
\centering
\vspace{-5mm}
\includegraphics[width=0.9\textwidth]{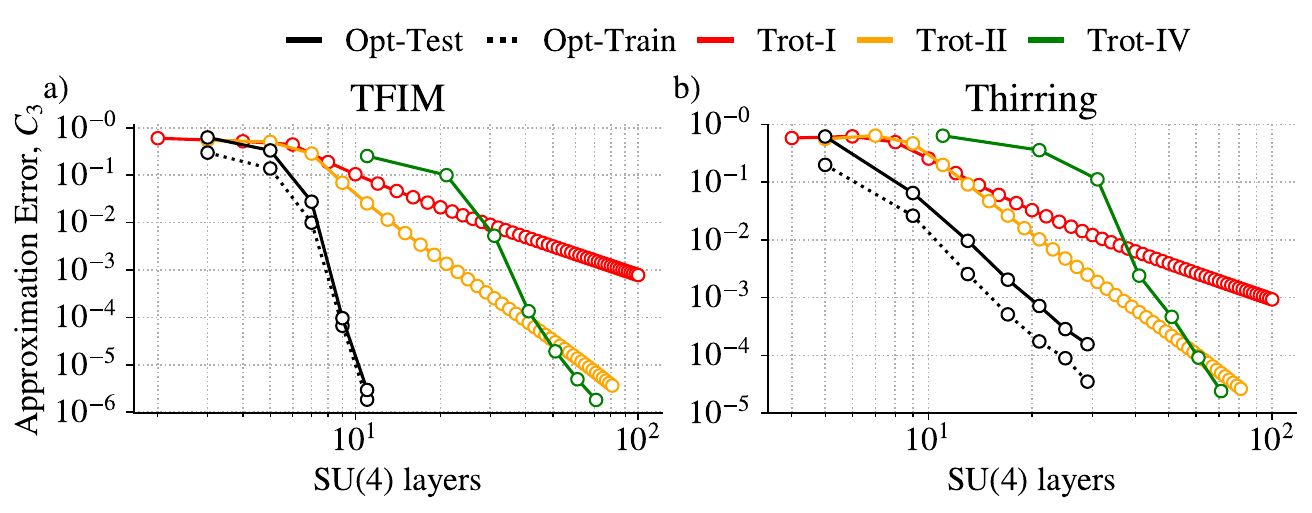}
\vspace{-3mm}
\caption{
\textbf{Circuit Compilation: Unitary Compression. }
We use the iMPS-based circuit compilation to find a short depth circuit to realize the evolution operator, $e^{-iHt}$, for two Hamiltonians. We optimize the cost function (Eq.~\eqref{eq:Cost_Uni}) for a set of 4 training states, the converged value is shown by the black dotted line.
We evaluate the progress by computing the cost function with a testing set of 100 unseen states, shown by the bold black line.
We compare these errors against equivalent depth Trotterizations, with the red/orange/green lines respectively corresponding to 1st/2nd/4th order Trotterizations. a) The target Hamiltonian is the Transverse-Field Ising model (Eq.~\eqref{eq:Ham_TFIM}), with a transverse-field strength of $h=1$. The total time evolution for this operator is $t=3.0$.
b) The target Hamiltonian is the Thirring model (Eq.~\eqref{eq:Ham_Thirring}). The total time evolution for this operator is $t=8.0$.
}
\label{fig:UnitaryCompress}
\end{figure*}

\subsection{Unitary Compression}\label{sec:unitary_compression}

In our third demonstration, we target the optimization of a circuit to realize the evolution operator $e^{-iHt}$, with a smaller error than equivalent depth circuits derived from Trotterization. Compared to the preparation of a time-evolved state in Sec.~\ref{sec:state_evolution}, here we target the more challenging task of approximating the full unitary of the propagator.

While we are now compiling a circuit to realize a unitary, rather than prepare a target state, we can continue working with iMPS. Recent results described a unitary compilation inspired by quantum machine learning~\cite{caro2022outofdistribution}, that can be faithfully performed only using the action of the unitary on a set of low-entangled input states. This is a form of ``Out-of-Distribution Generalization'', as the training set is restricted to low-entangled states, however the compilation is analytically guaranteed to generalize across the entire Hilbert space, even to highly entangled states. These low-entangled states are computationally efficient to operate on with tensor network methods, and a unitary compilation based on finite MPS has been studied~\cite{zhang2024scalable}. Here we extend this approach to infinite systems using iMPS. 

A unitary compilation based on applications of the circuit onto low-entangled states motivates the following cost function
\begin{equation}\label{eq:Cost_Uni}
    C_{3}(\thv) = 1 - \frac{1}{N} \sum_{k=1}^{N}\big|\lambda_0(V(\thv)  |\Phi^\chi_k \rangle, e^{-iHt}|\Phi^\chi_k \rangle)\big|^2.
\end{equation}
In the above definition, the state $|\Phi^{\chi}\rangle$ denotes a random iMPS with bond dimension $\chi$, and our training set is composed of $N$ of these random states. 
Previous work approached this compilation using unentangled input states e.g. $\chi_{\text{Train}}=1$. However, here we use $\chi_{\text{Train}}=2$ as this larger set of states has improved generalization properties, with a minimal increase in computational cost. 
The set of states $\{e^{-iHt}|\Phi^\chi_k \rangle \}_{k=1}^{N}$ are computed using iTEBD, where the evolution operator is represented with a high-accuracy Trotterization, converged in Trotter error.

As we exploit the translation invariance symmetry, our variational circuits have few parameters, due to the lack of scaling with respect to system size.
Therefore, $\mathcal{O}(L)$ variational parameters fully express an $L$-layer circuit. Consequently, using the results of Ref.~\cite{caro2022outofdistribution}, only $\mathcal{O}(L \ \mathrm{log}(L))$ training states are required for a faithful compilation. For our numerical analysis, we set $N_{\mathrm{Train}}=4$, which we find to be sufficient for all circuit depths tested. This compilation is successfully demonstrated on up to $L=29$ $SU(4)$ layers (see Fig.~\ref{fig:UnitaryCompress} b)) with 435 variational parameters. This highlights the observation that far fewer training states are required than the analytic upper-bound suggests.
To verify our compilation is successfully generalizing, we compute the cost function using an unseen set of $N_{\mathrm{Test}}=100$ random testing states of $\chi_\text{Test}=2$  , and confirm that the training and testing cost are minimized simultaneously. 
We further provide a numerical study of the rate of generalization as a function of the number of training states and their bond dimension in Appendix ~\ref{sec:Generalization}. Larger bond dimension training states have a corresponding increase in training time, however even the smallest increase from $\chi_\text{Train}=1$ to 2 finds significantly improved trainability, with the onset of generalization (the simultaneous decrease of the testing cost with the training cost) found at only 2 training states. This highlights the training efficiency achieved due to the small number of variational parameters after exploiting translation invariance.

The results of this demonstration are shown in Fig.~\ref{fig:UnitaryCompress}. Again we compare against 1st/2nd/4th-order Trotterized circuits across a range of Trotter numbers, shown by the red/orange/green curves respectively. For the target TFIM (Thirring) Hamiltonian, we optimize increasing depth circuits against Eq.~\eqref{eq:Cost_Uni} until convergence. The final training and testing costs are shown by the black curve. The deepest circuits tested of 11 (29) $SU(4)$ layers achieve over 3 (1) orders of magnitude reduction in error, compared to the best equivalent depth Trotterization.

\begin{figure*}[t!]
\centering
\vspace{-5mm}
\includegraphics[width=0.7\textwidth]{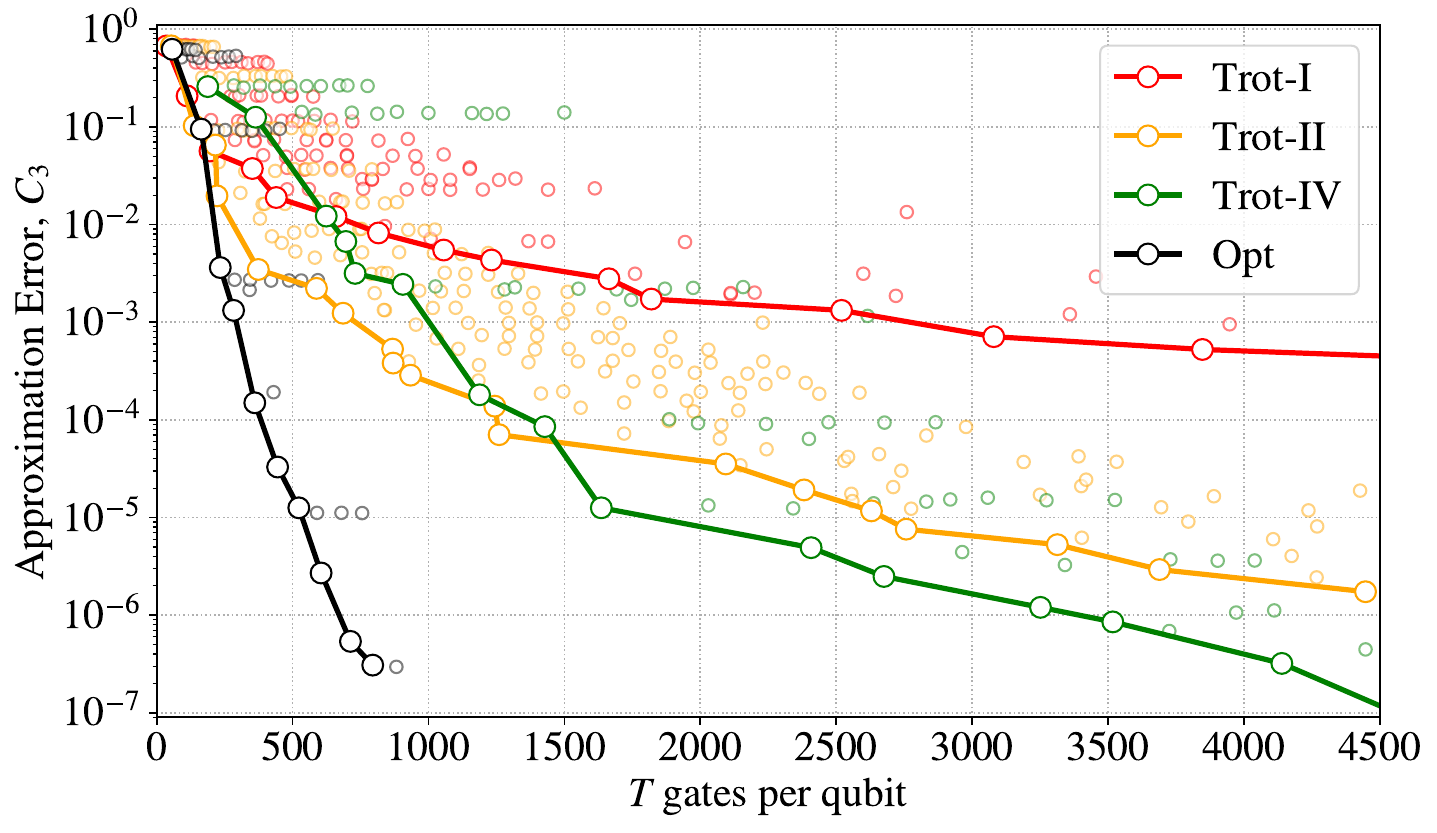}
\vspace{-3mm}
\caption{
\textbf{Advantage in $T$-count Reduction} Data showing the compilation of circuits expressing the unitary $e^{-iHt}$, for the TFIM Hamiltonian with $t=2.0$, into the Clifford+$T$ gate set. Trotterized circuits are compared to our iMPS-optimized circuits. 
For all circuits considered, we first express them in the Clifford+$R_z$ gate set, then decompose each $R_z$ gates to the Clifford+$T$ gate set using the Ross-Selinger algorithm to a range of precisions $\epsilon$. 
For the circuits derived from 1st/2nd/4th-order Trotterization, on Trotter numbers up to 100/50/20, we plot the corresponding $T$-count per qubit and approximation error ($C_3$), shown by the red, orange green colored circles respectively. We highlight the boundary of Trotter circuits with lowest $T$-counts by the bold lines.
The iMPS-optimized circuits, shown by the black circles, are found to need far fewer $T$ gates to reach a target error rate, with up to a $5.2\times$ reduction in $T$-count.
}
\label{fig:Clifford+T}
\end{figure*}
\subsection{$T$-Count Reduction}\label{sec:T-Count}

The unitary compression in the previous section showed our variational circuits can express the propagator $e^{-iHt}$ to a target approximation error with a smaller 2-qubit gate count compared to all equivalent depth Trotterizations. While this provides a significant advantage for quantum simulations on NISQ devices where 2-qubit gates are the dominant source of errors, these compressed circuits do not immediately imply an advantage for an error-corrected quantum computer.

Fault-tolerant algorithms are typically expressed in a gate set of Clifford gates, plus a non-Clifford gate providing universality. The most common is the Clifford+$T$ gate set, composed of the gates $\{H,S,CNOT,T\}$. Here, the $T$ gate is the non-Clifford gate that is expensive on fault-tolerant quantum hardware, as it requires magic state distillation~\cite{bravyi2012magic} to produce. In comparison, the Clifford gates are ``free'' or at least vastly cheaper.

We now consider the following problem: to represent the unitary $e^{-iHt}$ with a circuit expressed in the Clifford+$T$ gate set, can a target approximation error be reached with fewer $T$ gates per qubit when using iMPS compiled circuits, compared to all equivalent $T$-count Trotterizations? 

First, we describe the decomposition of the circuits derived from Trotterizations. For the target unitary $e^{-iHt}$, our Hamiltonian will be the TFIM with $h=1.0$, and $t=2.0$.
As done in Sec.~\ref{sec:unitary_compression}, the training states are prepared using iTEBD with a high accuracy Trotterization converged in time step.
We compare circuits derived from 1st/2nd/4th order Trotterizations across a large range of Trotter numbers. The Hamiltonian is split into $H = H_{ZZ} + H_X$; these contain the 2-body and 1-body interactions respectively, and are composed of entirely commuting terms. 
This allows the 3 Trotterization orders to be all constructed from alternating layers of exponentials of $H_{ZZ}$ and $H_X$. These circuits admit a simple decomposition to the Clifford+$R_z$ gate set, as the exponential of each term in the Hamiltonian requires one $R_z$ gate.

Then, we decompose each $R_z$ gate into the Clifford+$T$ gate set using the efficient Ross-Selinger algorithm~\cite{ross2014optimal}. The input is the angle $\theta$ defining the rotation of the $R_z(\theta)$ gate, and an $\epsilon$ which controls the precision the $R_z$ gate is compiled to with $||U - R_z(\theta)||\leq\epsilon$, where $U$ is the approximate gate with an explicit Clifford+$T$ decomposition. The higher the precision, the more $T$ gates are required, with a scaling of $\mathcal{O}(\text{log}(\frac{1}{\epsilon}))$. 

For every circuit, across the range of Trotter order and numbers, we apply the Ross-Selinger algorithm to the $R_z$ gates with increasing precision. Specifically, for every $\epsilon \in \{10^{-2.0},10^{-2.5},\ldots,10^{-7.0}\}$, we decompose each $R_z$ in the circuit to the chosen precision. We use that information to determine the total number of $T$ gates per qubit for that decomposed circuit. These new approximate circuits expressed in the Clifford+$T$ gate set are then used to compute the cost function in Eq.~\eqref{eq:Cost_Uni}.

We perform a similar decomposition for circuits produced by our iMPS-based optimization. 
As we seek circuits with low $T$-count, we constrain our circuits to contain fewer $R_z$ gates during the optimization to aid this. Rather than constructing our translation invariant layers with general $SU(4)$ gates, needing up to 15 $R_z$ gates to express, we replace this with an explicit decomposition to few $R_z$ gates 
$$
U(\theta_1, \theta_2, \theta_3) = R_{zz}(\theta_1) (R_x(\theta_2) \otimes R_x(\theta_3)).
$$
This two-qubit gate naturally encodes a first-order Trotterization of the target unitary, which serves as the initialization in our approach, with Trotter numbers ranging from 1 to 6. Then, during the optimization we simply change the angles of the rotation gates to minimize the approximation error, Eq.~\eqref{eq:Cost_Uni}, until convergence. As done for the Trotterized circuits, we again decompose all $R_z$ gates in a circuit across the same range of precisions by the Ross-Selinger algorithm to approximate our iMPS-optimized circuits in the Clifford+$T$ gate set, and recompute the cost function with the approximate circuits.

To recap, we have a collection of circuits derived from a broad range of order and depth Trotterizations of $e^{-iHt}$, and iMPS-optimized circuits. For each circuit, we decompose the $R_z$ gates to the Clifford+$T$ gate set to a range of precisions, and then compute the approximation error of these circuits compared to $e^{-iHt}$. This gives us a large dataset of the number of $T$ gates per qubit for the circuit, and the corresponding error. The dataset of Trotter (iMPS-optimized) circuits are represented by the colored (black) points in Fig.~\ref{fig:Clifford+T}. As a visual aid, we highlight the boundary circuits for each group with the lowest $T$-counts by the bold lines.

Our iMPS-based optimization is found to generate circuits in the Clifford+$T$ gate set with significantly fewer $T$ gates per qubit compared to circuits derived from Trotterizations. This can be highlighted by the iMPS-optimized circuit with 795 $T$ gates per qubit and corresponding approximation error $3.082 \cdot 10^{-7}$, compared to the data point for the best equivalent circuit derived from Trotterization, a 4th-order Trotterization with 4140 $T$ gates per qubit and approximation error $3.208\cdot 10^{-7}$, resulting in a $5.21\times$ reduction in $T$-count.

While tensor network circuit compilations have previously motivated the benefit for NISQ computers in minimizing CNOT count, here we give positive evidence this class of algorithms can also provide benefit to quantum simulations on fault-tolerant quantum computers, where reducing non-Clifford gate counts is the primary concern.

\begin{figure}[t!]
\centering
\vspace{-5mm}
\includegraphics[width=\columnwidth]{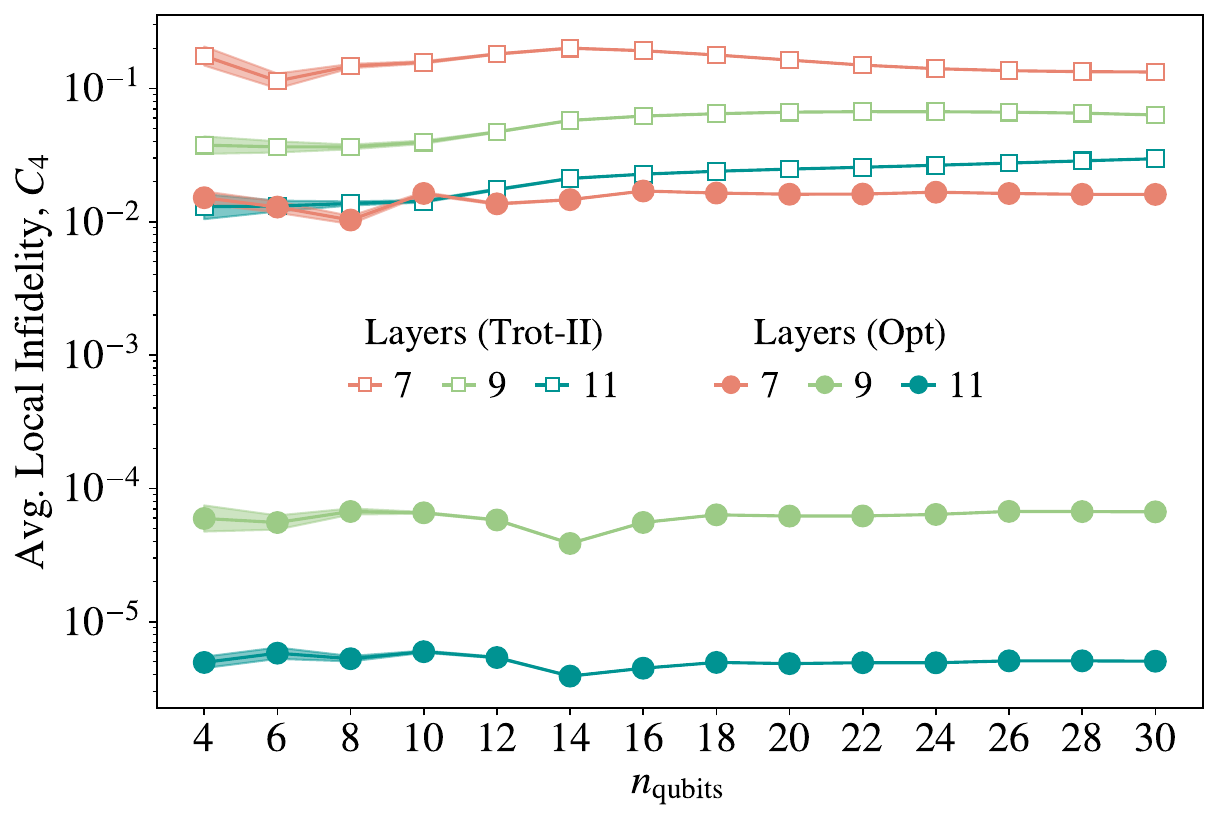}
\vspace{-3mm}
\caption{
\textbf{Infinite to Finite}
We validate our iMPS-compiled circuits aid simulation on finite numbers of qubits, using exact statevector simulation. To compare against the exact time evolution operator of the TFIM Hamiltonian, we compute the quantity Eq.~\eqref{eq:Cost_Finite} for the optimized circuits presented in Sec.~\ref{sec:unitary_compression}, compared to 2nd-order Trotterizations of the same depth, shown by the colored circles and the white squares respectively. The shaded region is the standard deviation of the $N=8$ local infidelities computed. For all qubit numbers tested, the optimized circuits have lower errors compared to the equivalent depth Trotterization, with an greater decrease in error as the circuit depth increases.
}
\label{fig:InfiniteToFinite}
\end{figure}

\subsection{Infinite to Finite}\label{sec:InfiniteToFinit}

We have presented a method for expressing and optimizing translation invariant 1D circuits in the thermodynamic limit. As quantum hardware advances, we expect to have access to larger qubit counts.
Therefore, it is important to verify our circuits trained in the infinite limit realize the claimed performance when used for a finite simulation, where finite-size effects may interfere.
Here it is most natural to study a 1D system with periodic boundary conditions (PBC). We use exact state vector simulation for the following calculations.

We perform the validation for the circuits optimized in the infinite limit in Sec.~\ref{sec:unitary_compression} to compress the evolution operator of $e^{-iHt}$ of the TFIM Hamiltonian. The optimized gates defining translation invariant circuit layers are used for realizing the time evolution operator on the finite 1D Hamiltonian with PBC, and are compared against circuits derived from Trotterization.

In comparison to the target unitary, we apply the optimized circuits and Trotterized circuits to Haar random states, denoted $|\mathcal{R}\rangle$, and compute inner products between these, resulting in the following cost function
\begin{equation}\label{eq:Cost_Finite}
C_4(V) = 1 - \frac{1}{N} \sum_{k=1}^{N} \big| \langle\mathcal{R}_k |V^{\dagger}e^{-iHt} | \mathcal{R}_k\rangle \big|^{2/{n_{\text{qubits}}}}.
\end{equation}
These inner products enable a measure of distance between the unitaries due to the approximation
$$1 - |\langle\mathcal{R}|V^{\dagger}U|\mathcal{R}\rangle|^2 \approx 1 - \frac{1}{2^{2n}}|\text{Tr}(V^{\dagger}U)|^2$$ with an error that decreases exponentially in the number of qubits, $n$~\cite{jin2021random}. This has a benefit in memory usage because the calculation can be performed through manipulation of length $2^n$ state vectors, rather than dimension $2^n \times 2^n$ unitary matrices. We modify the calculation to compute a local infidelity for a more similar measure to our earlier numerical results. We average over $N=8$ Haar random states. 

Fig.~\ref{fig:InfiniteToFinite} shows the result of this calculation. The optimized gates found in the infinite limit are used to create 1D translation invariant circuits to realize the time evolution operators for the $n$-qubit TFIM Hamiltonian with PBC, represented by the data points with colored markers. 
We compare against equivalent depth 2nd-order Trotterizations, as this Trotter order produces the lowest approximation errors for the circuit depths studied.
The 3 colors correspond to increasing circuit depths of $\{7,9,11\}$ layers of $SU(4)$ gates. The shaded region corresponds to the standard deviation due to the $N=8$ random states averaged over.
The x-axis shows the number of qubits $n_{\text{qubits}}$, the corresponding cost function value for these circuits Eq.~\eqref{eq:Cost_Finite} is displayed on the y-axis.
The large separation in error from the circuits derived from Trotterizations again shows the benefit gained from our iMPS-based optimization, with over 3 orders of magnitude reduction in error for the circuits of 11 $SU(4)$ layers.

These results highlight that while our optimization is designed to be faithful in the infinite limit, these benefits do persist when mapped onto the finite qubit systems. 
In practice, the natural use of our iMPS-based compilation to support quantum simulation on today's quantum computers is in simulating systems with 1D PBC.
We highlight that MPS simulations on systems with PBC is significantly more challenging, with an increase in bond dimension from $\chi$ to roughly $\chi^2$ compared to an equivalent simulation of open boundary conditions (OBC) due to the extra interaction between the ends of the MPS. Our method bypasses this constraint for compiling circuits with PBC, compared to other methods based on finite MPS.

While our circuits naturally realize translation invariant gate layers for PBC systems, they will likely  have utility for large OBC simulations. Here, the iMPS-optimized gates improve the simulation of the bulk of the system, and a re-optimization of the circuit with a finite MPS compilation~\cite{gibbs2024deep} adjusts for the relatively small boundary effects. A similar approach was motivated in ~\cite{mansuroglu2023variational, sokolov2025bang}.

\section{Discussion}\label{sec:discussion}

There is growing research into the use of tensor networks to support quantum simulations by reducing circuit resources. 
Material science is a key application for Hamiltonian simulation, where properties in the thermodynamic limit are of interest.
By exploiting the translation invariance symmetry, we develop a circuit compilation algorithm based upon iMPS with significantly simplified optimization complexity.
The power of tensor network enables a global optimization, rather than exploiting local properties.
Our algorithm learns a circuit to realize the unit-cell of a target state or unitary. The computational cost scales with the size of the unit cell, rather than the full system size. Far fewer variational parameters can express the circuit, similarly reducing the required number of training states. This unit cell creates translation invariant layers, enabling quantum simulations of arbitrary but finite-size systems on available quantum hardware.

Our iMPS-optimized circuits can prepare 1D ground states to high accuracy, and in simulations involving time-evolution are found to be more efficient in 2-qubit gate depth than all Trotterizations tested, improving the amenability to simulations on near-term hardware. For fault-tolerant quantum computers, non-Clifford gates are instead the expensive operation to be minimized. In compilations of our iMPS-optimized circuits to realize the unitary $e^{-iHt}$, after decomposing into the Clifford+$T$ gate set we similarly find an advantage, with up to a $5.2\times$ reduction in $T$-count.

Future work could explore extending these methods to be scalably applied to infinite 2D systems, and applied to more realistic Hamiltonians derived from material science. Our Clifford+$T$ decomposition was performed after the optimization had completed; further studies constraining $T$-count during the optimization, and using more sophisticated decomposition methods~\cite{hao2025reducing}, would likely find an even greater improvement.

\begin{acknowledgments}

J.G. was supported with funding and computational resources provided by AWE.
L.C. was supported by Laboratory Directed Research and Development program of Los Alamos National Laboratory under project number 20230049DR as well as by the U.S. Department of Energy, Office of Science, Office of Advanced Scientific Computing Research under Contract No. DE-AC05-00OR22725 through the Accelerated Research in Quantum Computing Program MACH-Q project.

\end{acknowledgments}

\newpage

\bibliography{quantum.bib}

\clearpage 
\appendix

\setcounter{page}{1}
\renewcommand\thefigure{\thesection\arabic{figure}}
\setcounter{figure}{0} 

\onecolumngrid

\begin{center}
\large{ Supplementary Material for \\ ``Learning Circuits with Infinite Tensor Networks''
}
\end{center}

\section{Generalization in Unitary Compilation}\label{sec:Generalization}

Here we study the impact on generalization performance due to the number of training states and their bond dimension. 

We perform a range of optimizations to compile the unitary $e^{-iHt}$ for the TFIM Hamiltonian (with $h=1$, $t=2$). The variational circuit has 11 layers of $SU(4)$ gates, resulting in 165 variational parameters in total. 
During the optimization, Eq.~\eqref{eq:Cost_Uni} is minimized, with $N_{\text{Train}}$ training states of random iMPS with bond dimension of both $\chi_{\text{Train}}=1$ and $\chi_{\text{Train}}=2$. The testing cost is computed with 100 unseen $\chi_{\text{Test}}=2$ random iMPS.
The dotted line shows the training cost and testing cost being exactly equal, which in this context corresponds to perfect generalization during training. 
For $N_\text{Train}=1$, generalization is not observed, signified by the plateau in testing cost while the training cost decreases below $10^{-6}$. 

As expected, the more training states are used, improved generalization is found in the reduced difference between training and testing costs. The generalization is found to increase significantly faster using the $\chi_{\text{Train}}=2$ random training states, where at $N_\text{Train}=3$ there is already a tight relationship between the training and testing costs.

\begin{figure}[h]
\centering
\includegraphics[width=0.6\textwidth]{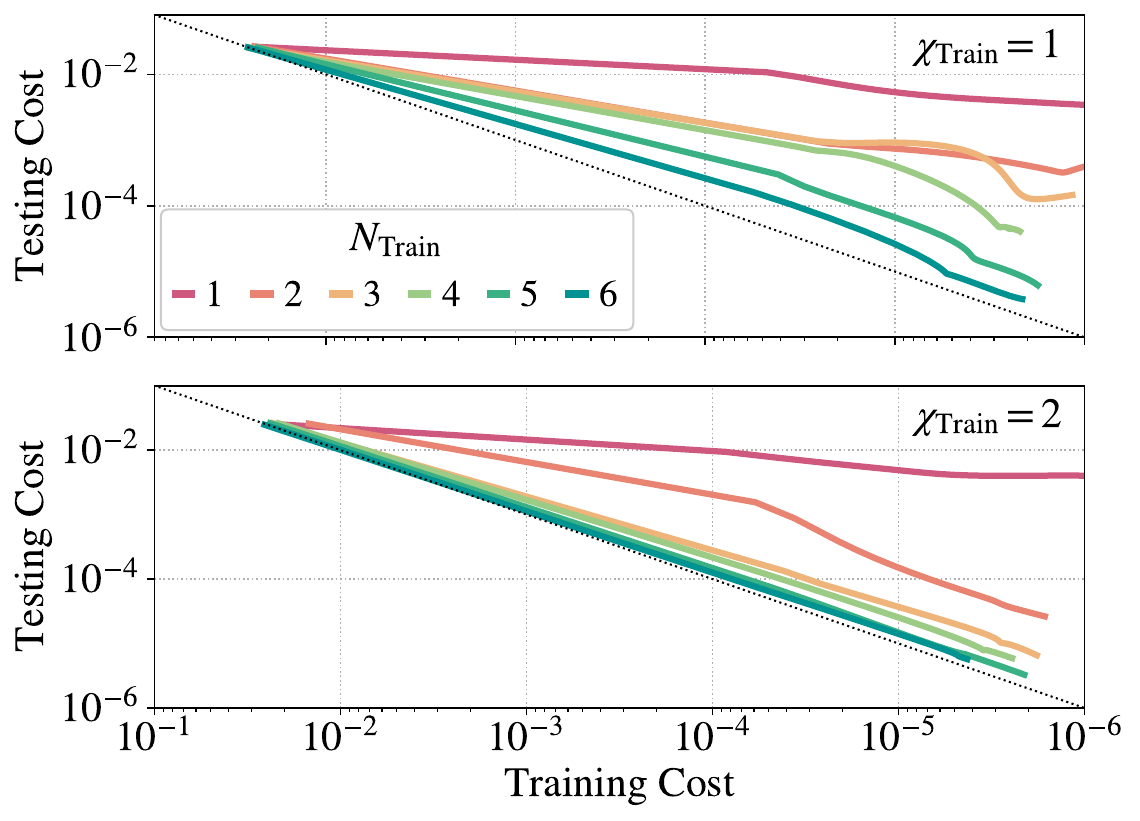}
\caption{
\textbf{Generalization with increasing size training dataset. }
The rate of convergence in generalization is studied for increasing size training sets. $\chi_\text{Train}$ is the bond dimension of the random iMPS used as training state in Eq.~\eqref{eq:Cost_Uni}. }
\label{fig:Generalization}
\end{figure}

\clearpage

\end{document}